%% file: Lattice_2014_proceedings.tex
\documentclass{PoS}

\title{\vspace{-1cm}
	{\small \normalfont \hfill DESY 14-248\hspace{0.3cm} HU-EP-14/67\hspace{0.3cm} SFB/CPP-14-105\\
}
\vspace{1cm} Phase structure and Higgs boson mass in a Higgs-Yukawa model with a dimension-6 operator}

\ShortTitle{Phase structure and Higgs boson mass in a Higgs-Yukawa model with a dimension-6 operator}

\author{David Y.-J.\ Chu\\
       National Chiao Tung University, Hsinchu, Taiwan \\
       E-mail: \email{ren1072.ep99@g2.nctu.edu.tw}
	 }

\author{Karl Jansen\\
       NIC, Desy Zeuthen, Germany\\
       E-mail: \email{karl.jansen@desy.de}
	 }

\author{Bastian Knippschild \\
       HISKP, Bonn, Germany\\
       E-mail: \email{knippschild@hiskp.uni-bonn.de}
	 }

\author{C.-J. David Lin \\
       National Chiao Tung University, Hsinchu, Taiwan\\
       E-mail: \email{dlin@mail.nctu.edu.tw}
	 }

\author{Kei-Ichi Nagai \\
       KMI, Nagoya, Japan\\
       E-mail: \email{keiichi.nagai@kmi.nagoya-u.ac.jp}
	 }
	
\author{\speaker{Attila Nagy}%
        \\
        Humboldt University Berlin; NIC, Desy Zeuthen, Germany\\
        E-mail: \email{nagy@physik.hu-berlin.de}
	  }

\abstract{We investigate the impact of a $\lambda_6 \varphi^6$ term included in a
chirally invariant lattice Higgs-Yukawa model. Such a term could emerge from BSM physics at some larger energy scale.
We map out the phase structure of the Higgs-Yukawa model with positive
$\lambda_6$ and negative quartic self coupling of the scalar fields. To this
end, we evaluate the constraint effective potential in lattice perturbation theory and also determine the magnetization of the model
via numerical simulations  which allow us to reach also non-perturbative values of the couplings. As a result, we find a complex phase structure with 
first and second order phase transitions identified through the magnetization.
Further we analyze the effect of such a $\varphi^6$ term on the lower Higgs boson
mass bound to see, whether the standard model lower mass bound
can be altered.}

\FullConference{The 32nd International Symposium on Lattice Field Theory,\\
		23-28 June, 2014\\
		Columbia University New York, NY}

\usepackage[font=footnotesize,labelfont=bf]{caption}

\usepackage{amssymb,amsmath,textcomp}
\usepackage{graphicx} 
\usepackage[font=scriptsize,labelfont=bf]{subfig}	
\usepackage{slashed}

\newcommand{\GeV}{\text{GeV}}
\begin{document}
\vspace{-3 mm}
\section{Introduction}
\vspace{-4 mm}
\input{Lattice_2014_intro.tex}

\vspace{-3 mm}
\section{The Higgs-Yukawa model on the lattice}
\vspace{-4 mm}
\input{Lattice_2014_HY.tex}

\vspace{-3 mm}
\section{Results on the phase structure}
\vspace{-4 mm}
\input{Lattice_2014_results_phase.tex}

\vspace{-3 mm}
\section{Results on the Higgs boson mass}
\vspace{-4 mm}
\input{Lattice_2014_results_mass.tex}

\vspace{-3 mm}
\section{Summary and conclusion}
\vspace{-4 mm}
\input{Lattice_2014_summary.tex}

\vspace{-3 mm}
\section{Acknowledgements}
\vspace{-4 mm}
\input{Lattice_2014_thanks.tex}

\bibliographystyle{unsrt}
\bibliography{Lattice_2014_references.bib}
%

\end{document}

%% file: Lattice_2014_intro.tex
With the discovery of the Higgs boson all ingredients in the standard model (SM) of particle physics have been completed in a sense, that it indeed 
describes particle interactions in the regime of energies that can presently be probed experimentally. There are however phenomena that cannot be explained by 
the standard model, like dark matter, dark energy and the incorporation of gravity in the framework of the other three known fundamental forces. 
Thus, the SM can account for all phenomena and needs to be replaced at some yet unknown 
energy. 
In~\cite{Degrassi:2012ry} this question has been analyzed through the stability of the electro-weak vacuum  
and a lower bound for the Higgs boson mass was found that is required to obtain a 
fully stable vacuum. 
This bound was found to be at~$m_H > 129.4 \pm 1.8~\GeV$.
The main uncertainties originates from the top quark mass  and the strong coupling,
determined at the mass of the Z-boson $\alpha_s(m_Z)$. 
The mass bound 
was obtained by running all SM couplings up to the Planck scale and requiring that the 
quartic self coupling of the scalar 
doublet $\lambda$ remains positive. 
A Higgs boson with a mass slightly below the bound derived in~\cite{Degrassi:2012ry} may yield a 
meta stable 
vacuum: The system can remain in a local minimum of energy with a non-vanishing probability 
to tunnel into the global vacuum with lower energy. Depending on the parameters, such 
a meta stable state can have a mean life time that exceeds the life time of the universe, 
so that such a scenario does not have immediate consequences.   

In this work we pursue an approach that investigates how the standard model Higgs boson mass lowe bound 
can be altered by the inclusion of a higher dimensional operator in 
the electro-weak sector. Explicitly we add a $\lambda_6 \varphi^6$ term to the action. 
With $\lambda_6>0$, the action remains bounded from below even for negative 
quartic self coupling of the scalar fields. The inclusion of such a term may appear naturally if one 
considers the Higgs sector of the SM as a low energy effective theory 
obtained by integrating out degrees of freedom at some higher scale physics. 
For our investigations we use a chirally invariant lattice formulation of the Higgs-Yukawa 
model as a limit of the SM where only one family of quarks and the scalar doublet 
are considered. This model was already 
successfully used to determine the cutoff dependence of the Higgs boson mass bounds for the SM~\cite{Gerhold:2009ub, Gerhold:2010bh}, and 
to investigate the change on those bounds in case of a heavy fourth generation of quarks~\cite{Bulava:2013ep}.

Earlier works showed that in the Higgs-Yukawa model with $\lambda_6=0$ the phase transition of interest, i.e.\ the transition between a symmetric 
and a broken phase with small quartic couplings and yukawa couplings generating quarks with a mass of the order of the top quark mass, is of second 
order. This may change drastically with  
the inclusion of a $\lambda_6 \varphi^6$ term due to
the more complex structure of the bosonic potential. As we will see below,
depending on the choice of $\lambda$ and $\lambda_6$ there appear lines of first order 
transition separating the symmetric and the broken phases or even 
further transitions between different non-zero magnetizations. 

In this work we present results on the phase structure and the Higgs boson mass having 
$\lambda_6\neq0$. We study both aspects perturbatively by means 
of Lattice perturbation theory using a constraint effective potential (CEP) and non-perturbatively via lattice simulations.

%% file: Lattice_2014_HY.tex

The field content of the Higgs-Yukawa model in this work is given by a mass degenerate 
fermion doublet $\psi$ and the complex scalar doublet $\varphi$. The continuum formulation of the action of the Higgs-Yukawa model is given by: 
{\footnotesize
\begin{multline}\label{eq:action_continuum}
 S^{\text{cont}}[\bar{\psi}, \psi, \varphi] = \int d^4 x \left\{\frac{1}{2}\left(\partial_{\mu} \varphi \right)^{\dagger} \left(\partial^{\mu} 
\varphi 
\right)
			+  \frac{1}{2} m_0^2 \varphi^{\dagger} \varphi 
			+ \lambda \left(\varphi^{\dagger} \varphi \right)^2 
			+ \lambda_6 \left(\varphi^{\dagger} \varphi \right)^3 \right\} \\
			+\int d^4 x  \left\{\bar{t} \slashed \partial t + \bar{b} \slashed \partial b +
			y \left( \bar{\psi}_L \varphi\, {b}_{_R} + \bar{\psi}_L \tilde \varphi\, {t}_{_R} \right)
			+ h.c. \right\},
		\end{multline}}
with $\tilde \varphi = i\tau_2\phi^{*}$ and $\tau_2$ being the second Pauli matrix. 

For any details on the numerical implementation of the lattice simulations we refer to~\cite{Gerhold:2010wy}. 
We just mention that we use the
polynomial Hybrid Monte Carlo algorithm as the basic simulation tool.
For the discretization of the fermions we use the chirally invariant overlap operator. To set the scale 
we identify the vacuum expectation values ($vev$) of the scalar field with the phenomenological value of $vev \approx 246~\GeV$.

The perturbative results are obtained by using a constraint effective potential $U(\hat v)$ \cite{Fukuda:1974ey,O'Raifeartaigh:1986hi}. The basic idea 
is that the system is dominated by the zero mode of the scalar field, and therefore we can integrate out all non-zero modes. 
The $vev$ of the scalar field and the 
Higgs boson mass can be obtained by the global minimum and the curvature at the minimum of the potential, respectively:
{\footnotesize
\begin{equation}\label{eq:vev_and_mass_CEP}
  0=\frac{\text{d}U(\hat v)}{\text{d}\hat v} \Bigg|_{\hat v = vev} ,\qquad
 m_H^2 = \frac{\text{d}^2 U(\hat v)}{\text{d}\hat v^2} \Bigg|_{\hat v = vev}. 
\end{equation}}
We point out, that we explicitly keep the lattice regularization in the perturbative approach. Thus, 
we will work with only discrete momenta and we also use 
the overlap operator for the fermionic kernel.

We compare two expressions for the CEP which differ by the decomposition of the action into an 
interaction part that has to be expanded in powers of the couplings and a Gaussian part that can be integrated out. 
The first expression was already used in~\cite{Gerhold:2007gx,Gerhold:2009ub} for 
the Higgs-Yukawa model and is given to first order in $\lambda$ and $\lambda_6$ by:
{\footnotesize
\begin{multline}\label{eq:CEP_with_phi_6}
 U_1(\hat v) = U_f(\hat v) + \frac{m_0^2}{2} {\hat v}^2 +\lambda {\hat v}^4 + \lambda_6 {\hat v}^6 \\
                         + \lambda \cdot {\hat v}^2 \cdot 6(P_H+P_G)
                         + \lambda_6 \cdot \left( {\hat v}^2 \cdot ( 45 P_H^2 + 54 P_G P_H + 45 P_G^2)
                         + {\hat v}^4 \cdot ( 15 P_H + 9 P_G ) \right ).
\end{multline}}
In the propagator sums $P_{H/G}=\frac{1}{V}\sum_{p\neq 0}\frac{1}{\hat p ^2 + m_{H/G}^2}$ the masses are set ``by hand'' to zero for the Goldstone and 
to the Higgs boson mass obtained from eq.~\eqref{eq:vev_and_mass_CEP}. $U_f$ denotes the contribution from integrating out the fermions in 
the background of a constant field.

A second possibility to formulate the CEP is obtained, when taking zero-mode contributions from the self interactions of the scalar field into 
account, by  
integrating out the Gaussian part in the non-zero modes of the scalar field. In this approach, logarithmic dependence on $\hat v$ appear. Further the 
propagator sums and the first order contribution in the self couplings change, 
yielding:
{\footnotesize
\begin{align}\label{eq:CEP_with_phi_6_withFullBosDet}
 U_2(\hat v) & =  U_f(\hat v) + \frac{m_0^2}{2} {\hat v}^2 +\lambda {\hat v}^4 + \lambda_6 {\hat v}^6 
              + \frac{1}{2 V} \sum\limits_{p \neq 0} 
                 \log \left[  \left( \hat p^2 + m_0^2 + 12 \lambda \hat v^2 + 30 \lambda_6 \hat v^4 \right) \cdot  
                 \left( \hat p^2 + m_0^2 + 12 \lambda \hat v^2 + 30 \lambda_6 \hat v^4 \right)^3  \right]
         \nonumber \\
             & + \lambda \left ( 3 \, \tilde P_H^2 + 6 \, \tilde P_H \tilde P_G + 15 \, \tilde P_G^2 \right)   
                 + \lambda_6 \hat v^2 \left ( 45 \, \tilde P_H^2 + 54 \, \tilde P_H \tilde P_G + 45 \, \tilde P_G^2 \right) 
               + \lambda_6 \left( 15 \, \tilde P_H^3 + 27 \, \tilde P_H^2 \tilde P_G + 45 \, \tilde P_H \tilde P_G^2 + 105 \, \tilde P_G^3 \right),
			 \end{align}}
with 
{\footnotesize
\begin{equation}\label{eq:propagatorSum_withFullBosDet}
\tilde P_H = \frac{1}{V} \sum\limits_{p\neq 0} \frac{1}{ \hat p^2 + m_0^2 + 12 \hat v^2 \lambda + 30 \hat v^4 \lambda_6 }, \quad
\tilde P_G = \frac{1}{V} \sum\limits_{p\neq 0} \frac{1}{ \hat p^2 + m_0^2 + 4 \hat v^2 \lambda + 6 \hat v^4 \lambda_6 }.
\end{equation}}
This second method should be more precise but has a limited range of validity 
depending on the parameters: It fails, when the argument of the 
logarithms become negative leading in this case to a complex potential.

%% file: Lattice_2014_results_phase.tex
For our analysis we chose two values for $\lambda_6$, namely $\lambda_6=0.001$ and $\lambda_6=0.1$. 
Using the the tree-level relation $m_t= y \cdot vev$ we 
further fix the Yukawa coupling to yield the physical top quark mass.  
We work at 
several values for $\lambda$ and perform scans in $m_0^2$ or, equivalently, 
the scalar hopping parameter in the lattice action $\kappa$\footnote{
	The relation between $\kappa$ and $m_0^2$ is given by: ${m_0^2=\frac{1- 8 \lambda \kappa^2 - 8 \kappa}{\kappa}}$.}.
For the here performed phase structure study we use the $vev$ as an order 
parameter. In infinite volume the $vev$ is zero in the symmetric phase 
and non-zero in the broken phase. In finite volume the $vev$ never assumes an 
exactly zero value, so eventually an extrapolation to infinite volume must 
be carried out to unambiguously determine the phase structure of the system.

\begin{figure}[htb]
\centering
\subfloat[$\lambda_6=0.001$]{\includegraphics[width=0.31\linewidth]
{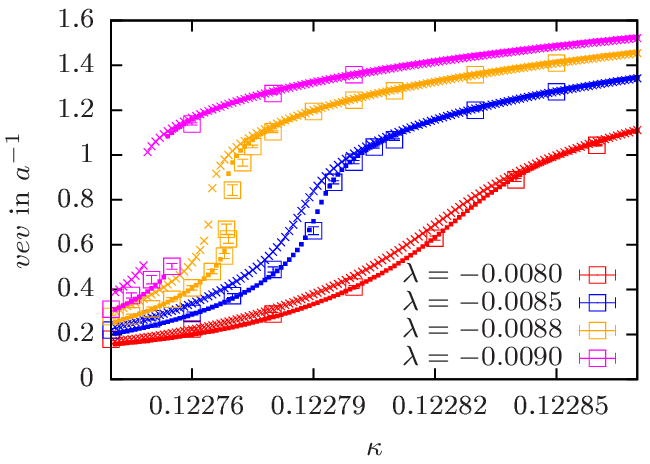}\label{fig:vev_vs_kappa_l6_0.001}}
~
\subfloat[$\lambda_6=0.1$]{\includegraphics[width=0.31\linewidth]
{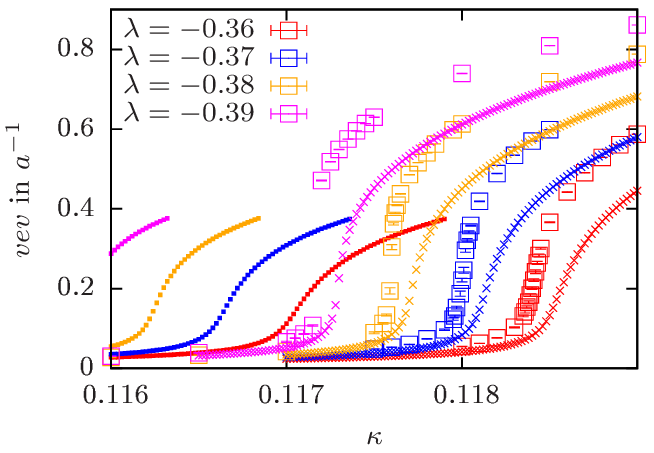}\label{fig:vev_vs_kappa_l6_0.1}}
~
\subfloat[$\lambda_6=0.001,\, \lambda=-0.0085$]{\includegraphics[width=0.31\linewidth]
{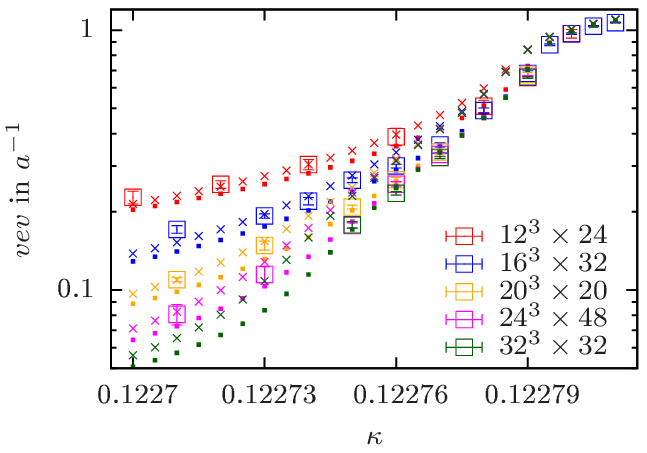}\label{fig:Vdep_l6_0.001_l_-0.0085}}
\caption{The first two plots show the $vev$ as a function of $\kappa$ for various values of $\lambda$ while $\lambda_6$ is kept fixed to 
$\lambda_6=0.001$ (a) and $\lambda_6=0.1$ (b) obtained on $16^3 \times 32$ lattices.
We show a comparison between data obtained from lattice simulations (open squares) with the CEP $U_1$ (crosses) and $U_2$ (dots).
In (c) the dependence on the volume is illustrated for $\lambda_6=0.001$ and $\lambda=-0.0085$.}
\label{fig:vev_vs_kappa}
\end{figure}


As a first step we compare results obtained from the CEP and lattice simulations in 
figure~\ref{fig:vev_vs_kappa_l6_0.001} and~\ref{fig:vev_vs_kappa_l6_0.1}
where we show the $vev$ against the 
hopping parameter $\kappa$ on a $16^3 \times 32$ lattice. We show for both values of $\lambda_6$ the results from lattice simulations 
and the analytical results from both CEPs in eqs.~\eqref{eq:CEP_with_phi_6} and~\eqref{eq:CEP_with_phi_6_withFullBosDet}. Both setups 
show qualitatively the same picture: With $\lambda_6>0$ and negative but small $\lambda$, 
the transition between a symmetric phase and a broken phase 
is continuous which suggests a second order transition, while for smaller 
$\lambda$ the order parameter shows 
a jump at some transition value 
indicating a transition of first order. 
For $\lambda_6=0.001$ both potentials describe the data very well, 
although the quantitative agreement is 
slightly better for the CEP $U_2$ in eq.~\eqref{eq:CEP_with_phi_6_withFullBosDet}. 
The potential $U_1$ in eq.~\eqref{eq:CEP_with_phi_6} also fails to exactly reproduce the 
first order phase transition for $\lambda=-0.0088$. However, if one decreases $\lambda$ further, also the potential $U_1$ shows a first order 
transition. 
For $\lambda_6=0.1$ the CEP $U_2$ does not reproduce the behavior of the simulation data. 
This is not surprising, since in this case analyzing the CEP $U_2$ in the region of parameter 
space close to the  
the phase transitions we meet the difficulty that the effective potential 
becomes complex.  
Qualitatively $U_1$ shows still a good agreement, but again 
the value of $\lambda$ where the transition turns first order is not reproduced exactly.
%
In~\ref{fig:Vdep_l6_0.001_l_-0.0085} we show the volume dependence of the $vev$ 
for fixed $\lambda_6=0.001$ and $\lambda=-0.0085$, comparing results obtained from 
both potentials and simulation data. We find, that the qualitative behavior of the simulation data is very well reproduced by both perturbative 
approaches.


Since the constraint effective potential describes the results from the simulations reasonably well, 
we will rely on the CEP to perform a more complete 
study of the phase structure with large volumes. Furthermore, we restrict ourselves to the potential $U_1$. 
The exact location of the phase transition in finite volume cannot easily be determined 
from the order parameter alone due to the fact that the $vev$ does not exactly go to zero. 
One possibility is to investigate the curvature of the 
potential at its minimum which is identical to the squared 
Higgs boson mass according to eq.~\eqref{eq:vev_and_mass_CEP}. The inverse curvature 
of the potential in the minimum is related to the  
susceptibility and the peak position of which can then be used for locating of the phase transition
point.  


\begin{figure}[htb]
\centering
\subfloat[$\lambda_6=0.001,\, \lambda=-0.0085$]{\includegraphics[width=0.31\linewidth]
{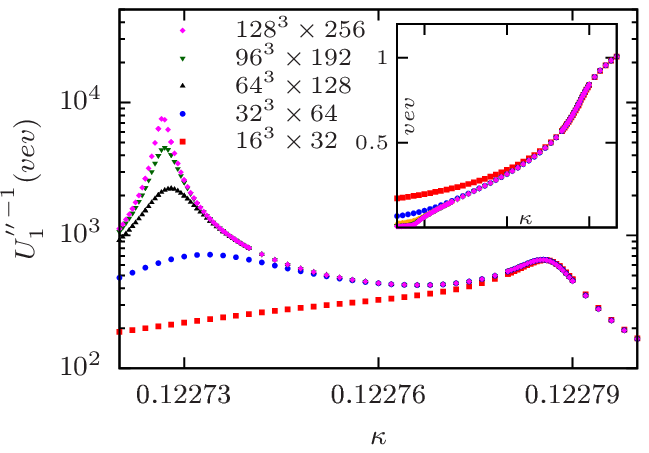}\label{fig:CEP_Vdep_mass_and_vev_l_-0.0085}}
~
\subfloat[$\lambda_6=0.001$]{\includegraphics[width=0.31\linewidth]
{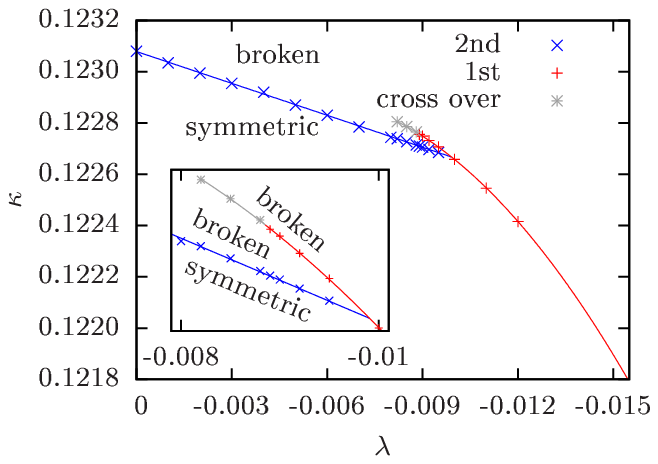}\label{fig:phaseStructure_l6_0.001}}
~
\subfloat[$\lambda_6=0.1$]{\includegraphics[width=0.31\linewidth]
{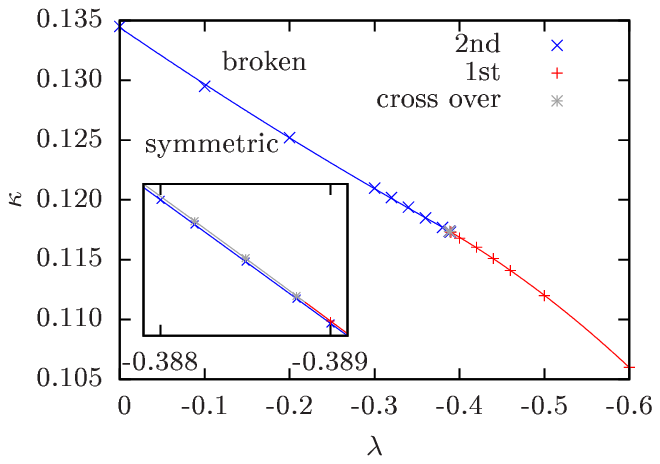}\label{fig:phaseStructure_l6_0.1}}
\caption{The first plot shows the $vev$ (inlet) and the inverse curvature at the minimum of the potential as a 
function of $\kappa$ for various volumes. The first peak is a good indicator for a second order transition, while the second peak 
without volume dependence may refer to a cross over. The other figures show the result of a more complete phase structure scan for $\lambda_6=0.001$ 
(middle) and $\lambda=0.1$ (right) obtained from the CEP $U_1$ \protect\eqref{eq:CEP_with_phi_6}. There are two phases - a broken and a symmetric one 
- separated by lines of first and second order phase transitions. Further there is a small region in 
parameter space, where there is also a first order transition between two broken phases (for 
$\lambda_6=0.001$ and $\lambda_6=0.1$). The lines connecting 
the data points are to guide the eye.
}
\label{fig:CEP_Vdep_mass_and_vev}
\end{figure}

In figure~\ref{fig:CEP_Vdep_mass_and_vev_l_-0.0085} an example plot is shown that 
illustrates the behavior of the order parameter and the inverse curvature 
of the potential at its minimum 
as a measure of the susceptibility for $\lambda_6=0.001$ and $\lambda=-0.0085$. The inverse curvature shows two peaks. The first peak is 
getting higher with increasing volume, a typical behavior of 
the peak of the susceptibility which is diverging at the critical coupling 
when the volume is increased to infinity. The second 
maximum does not increase with increasing volume which suggests that this is a 
crossover transition.

Performing such scans 
systematically, we obtain the phase diagrams in the $\kappa-\lambda$-plane for both fixed $\lambda_6=0.001$ in 
figure~\ref{fig:phaseStructure_l6_0.001} and 
$\lambda_6=0.1$ in figure~\ref{fig:phaseStructure_l6_0.1}. Qualitatively both diagrams show the same 
behavior: for $\lambda=0$ there is a single phase transition of second order 
separating the symmetric and the broken phases. If $\lambda$ is 
decreased, the transition point moves to smaller $\kappa$. 
A some point, a line of crossover transition appears in the broken phase which turns into 
a line of first order transition. At some critical point the line of second order 
transition joins the first order transition line. Beyond that, 
only a first order transition separates the symmetric and the broken phases. 
For $\lambda_6=0.1$ 
the region in the parameter space where the first order and the crossover 
transitions appear in the broken phase is very narrow. Given the fact that the 
agreement between the CEP and the simulation data for a value of $\lambda_6=0.1$ 
is not completely satisfactory, a test of the phase structure from 
numerical simulations would be very desirable. 


%% file: Lattice_2014_results_mass.tex
\begin{figure}[htb]
\centering
\subfloat[$\lambda_6=0.001$]{\includegraphics[width=0.31\linewidth]
{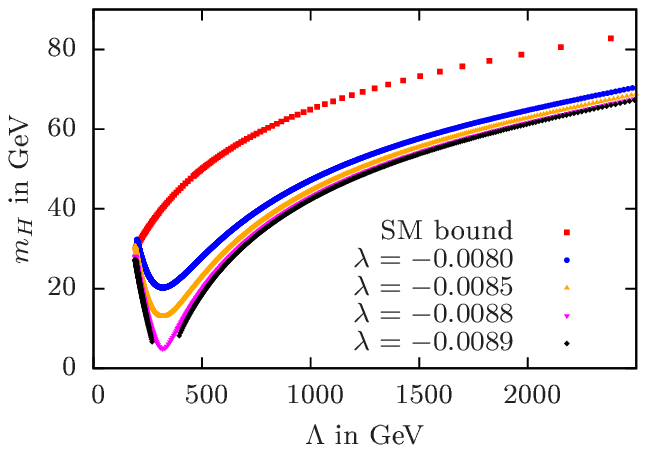}\label{fig:mass_vs_cutoff_l6_0.001}}
~
\subfloat[$\lambda_6=0.1$]{\includegraphics[width=0.31\linewidth]
{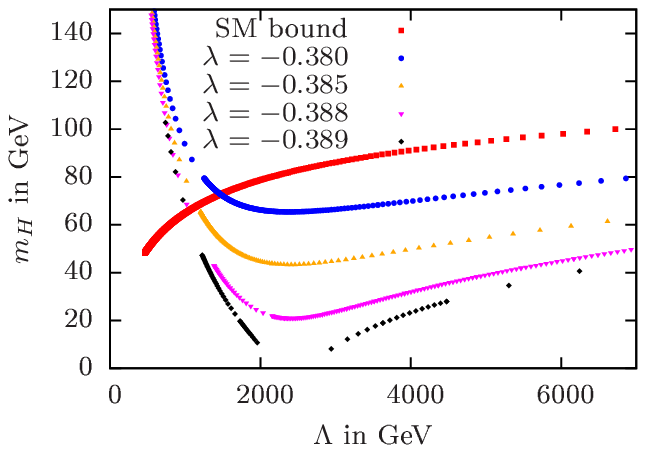}\label{fig:mass_vs_cutoff_l6_0.1}}
~
\subfloat[$\lambda_6=0.001$]{\includegraphics[width=0.31\linewidth]
{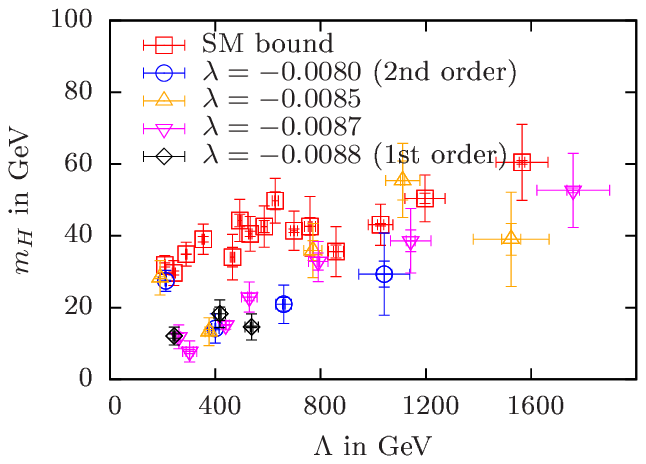}\label{fig:mass_vs_cutoff_sim_l6_0.001}}
\caption{Shown is the cutoff dependence of the Higgs boson mass obtained from the 
CEP according to eq.~\protect\eqref{eq:vev_and_mass_CEP} for $\lambda=0.001$ on a $64^3\times128$-lattice (left) and
$\lambda=0.1$ on a $192^3 \times 384$ (middle) and preliminary results from lattice simulations for $\lambda_6=0.001$. In all plots we also show the 
standard model lower mass bound ($\lambda_6=\lambda=0$).
}
\label{fig:mass_vs_cutoff}
\end{figure}

In this section, the influence of the $\lambda_6 \varphi^6$ term on the 
lower Higgs boson mass bound is discussed. Especially we look at the cutoff dependence of the 
lower Higgs boson mass. In figure.~\ref{fig:mass_vs_cutoff} we show the 
perturbative results obtained with the potential $U_1$ 
for both values of $\lambda_6$ used. We also show the result of the lower Higgs boson mass 
bound for the case of vanishing $\varphi^6$ term.  
All data with non-zero $\lambda$ and $\lambda_6$ show a similar behavior. 
For small cutoff values the mass increases substantially, a phenomenon which is not 
present for the case of vanishing couplings. This behavior is related to 
the mass shift 
from positive $\lambda_6$ values, which contribute significantly at small cutoff scales.
The Higgs boson mass then has a minimum at some given cutoff (for the data that show a second order 
transition) and increases as similar to the SM bound. Further we can see
that the Higgs boson mass gets smaller, when $\lambda$ is 
decreased. 
As a criterion, how low the quartic coupling can be driven, we use 
a value of $\lambda$ 
where the transition between the broken and the symmetric phase 
turns from second to first order.


For the case of $\lambda_6=0.001$ 
the lower Higgs boson 
mass bound is significantly decreased as also found in \cite{Gies:2013fua}. 
Thus, for such small values of $\lambda_6$ the here considered extension 
of the standard model with a $\varphi^6$-term is fully compatible 
with the 126~GeV Higgs boson mass.  
For $\lambda_6=0.1$ the situation changes since here the lower bound
meets the 126~GeV Higgs boson mass
at a cutoff of around 
800~GeV, which is still inside the scaling region. Thus, large 
values of $\lambda_6$ cannot be admitted as an extension of the standard 
model.  
For both values of $\lambda_6$ and at large cutoffs the influence of the 
$\varphi^6$ becomes negligible and we find the SM-like behavior 
of the lower Higgs boson mass bound as a function of the cutoff. 

Preliminary results on the cutoff dependence of the lower Higgs boson mass obtained
from non-perturbative lattice simulations can be found in
fig.~\ref{fig:mass_vs_cutoff_sim_l6_0.001} for a fixed value of $\lambda_6=0.001$.
We show data for values of $\lambda$ corresponding to ones used in
figure~\ref{fig:mass_vs_cutoff_l6_0.001}. Note that
for some of the intermediate $\lambda$ values the order of the phase transition
is still not clear.
Although the data for $\lambda \gtrsim -0.0087$, where there is still a second
order phasetransition, still have large error bars,
it is evident, that also
non-perturbatively the standard model lower bound can easily be
decreased
well below 126~GeV while keeping $m_H/\Lambda$ well below 1/2 and hence staying in the scaling region.

%% file: Lattice_2014_summary.tex
We studied the phase structure of a chirally invariant lattice Higgs-Yukawa model
-- allowing non-perturbative computations -- 
with the addition of a $\lambda_6 \varphi^6$ term as a simple model for 
an extension of the standard model. Having  
$\lambda_6>0$ allows to set 
the quartic coupling $\lambda<0$. We found good agreement between a lattice perturbation theory 
approach using the CEP and Monte Carlo simulations for the behavior of 
the vacuum expectation value as function of the bare mass.
A systematic study of phase transitions using the $vev$ and the susceptibility led 
to the phase diagrams shown in figures~\ref{fig:phaseStructure_l6_0.001} and~\ref{fig:phaseStructure_l6_0.1}. We found 
transitions of first and second order separating the symmetric and the 
broken phase as well as first order transitions separating two broken vacua 
indicating the possibility of meta stable states.
The appearance of first order phase transitions in the presence of
a $\varphi^6$-term can be very interesting 
for the case of a non-zero temperature. 
It might lead to a scenario where a simple addition of a $\varphi^6$-term
can provide a strong enough first order phase transitions to be 
compatible 
with electro-weak baryogenesis in the early universe \cite{Cohen:1993nk}.
Another aspect of the phase diagram in 
figure~\ref{fig:CEP_Vdep_mass_and_vev} 
is 
that at fixed bare Higgs boson mass one can move from 
a symmetric to a broken phase by only changing the value of the 
quartic coupling. 

Further we investigated the influence of the dimension-6 operator 
on the Higgs boson mass bound with respect to the question 
whether the addition of this operator can be 
compatible with a 126~GeV Higgs boson mass and whether the lower bound can be altered  
compared to the Higgs-Yukawa limit of the standard model (i.e.\ 
$\lambda_6=0$). 
We found that for the values of $\lambda_6$ considered 
here, at large values of the cutoff the lower bound can be significantly decreased 
before it becomes compatible with the case of $\lambda_6=0$ for increasing cutoffs.
However, for a larger value of $\lambda_6=0.1$
the lower Higgs boson mass bound meets the 126GeV Higgs boson mass already at a cutoff 
of about 800GeV. Thus, such a large value of $\lambda_6$ is excluded. 
We plan to determine 
a critical value of $\lambda_6$ from which on 
an extension of the standard model with a $\varphi^6$-term is not 
compatible anymore with the 126~GeV Higgs boson mass. This can in turn provide 
bounds on models beyond the SM that generate 
effectively such a term.

%% file: Lattice_2014_thanks.tex
The simulations have been performed at the  the PAX cluster at DESY-Zeuthen,
HPC facilities at National Chiao-Tung University and the cluster system $\varphi$
at KMI in Nagoya University. This work is supported by Taiwanese NSC via grants 100-2745-
M-002-002-ASP (Academic Summit Grant), 102-2112-M-009-MY3, and by the DFG through the DFG-project Mu932/4-4, 
and the JSPS Grant-in-Aid for Scientific Research (S) number 22224003.